\documentclass{IEEEtran}
\usepackage{cite}
\usepackage{amsmath,amssymb,amsfonts}
\usepackage{graphicx}
\usepackage[caption=false]{subfig}
\usepackage{multirow}
\usepackage{cite}
\usepackage{color}

\DeclareMathOperator*{\argmax}{arg\,max}

\begin{document}

\title{Decoding of Variable Length PLH Codes}

\author{Marco Morini, Alessandro Ugolini, and Giulio Colavolpe
\thanks{This work has been supported by MIUR under the PRIN Liquid\_Edge contract. This research benefits from the HPC (High Performance Computing) facility of the University of Parma, Italy.
The authors are with University of Parma, Department of Engineering and Architecture, and CNIT Research Unit, Parma, Italy (E-mail: marco.morini@unipr.it, alessandro.ugolini@unipr.it, giulio.colavolpe@unipr.it).}
}

\maketitle

\begin{abstract}
	In this paper, we address the problem of the decoding of variable length physical layer header (PLH) codes. We take as a case study the DVB-S2X standard, and, after selecting a suitable PLH code with variable codeword length, we propose two alternative noncoherent decoding strategies. The proposed decoding strategies allow to estimate the length of the transmitted PLH field, which is unknown at the receiver, jointly with the decoding of the PLH code. We demonstrate that it is possible to achieve an excellent decoding performance while significantly reducing the overhead due to the transmission of the PLH field with respect to a standard fixed-length PLH.
\end{abstract}


\section{Introduction}
\begin{figure}
	\centering
	\includegraphics[width=0.9\columnwidth]{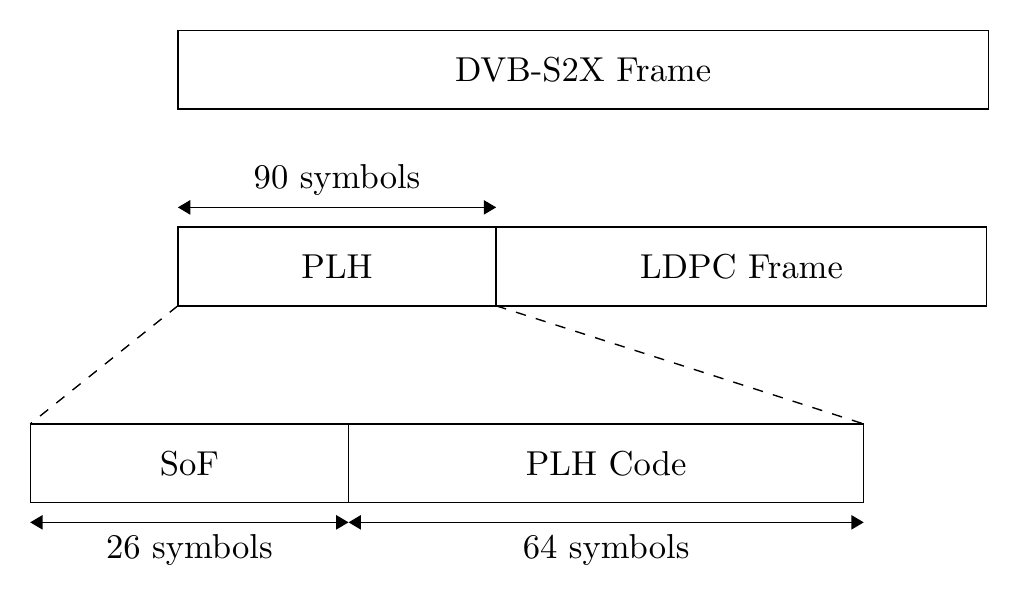}
	\caption{Frame structure for the DVB-S2X standard~\cite{DVB-S2X}.}
	\label{fig:pls_dvb}
\end{figure}

Many communication standards, for both broadcast and unicast applications, foresee the use of a physical layer header (PLH) field at the beginning of each transmitted frame. This is a control field which is usually exploited to mark the start of frames and for synchronization purposes. Moreover, PLH fields contain information on the content of the subsequent frame, like the indication of the modulation and coding formats (ModCod) adopted for the transmission. For example, the extension of the second generation of the digital video broadcasting via satellite (DVB-S2X) standard~\cite{DVB-S2X} adopts a $90$-bit PLH in each of the transmitted frames, which includes an invariant $26$-bit start-of-frame (SoF) sequence, followed by a $64$-bit PLH codeword, which contains the ModCod information. The PLH is then modulated by means of a $\pi/2-$binary phase-shift keying (BPSK).  DVB-S2X adopts constellations with sizes from $4$ to $256$, and a wide variety of coding rates to encode information by means of low-density parity-check (LDPC) codes, which define a set of $39$ ModCods (see~\cite{DVB-S2X}). The constellation size and the code rate represent the information that is encoded by means of a PLH code and enclosed in the PLH field at the beginning of each frame. Hence, it is mandatory to  successfully retrieve the information enclosed in the PLH before attempting the decoding of the transmitted information; in fact, the LDPC decoder needs to know the adopted modulation and coding formats, and this information can be achieved only if the PLH field is correctly decoded. A representation of the DVB-S2X frame structure is reported in Fig.~\ref{fig:pls_dvb}, where the division of the header in SoF and PLH code is highlighted.

Received signals are usually affected by unknown amplitude and phase offsets, which are detrimental for the decoding performance. These impairments can be estimated by means of pilot symbols, which are distributed in blocks at regular intervals in the frame. However, a variation of these effects over time can result in an imprecise pilot-based estimation, which would negatively affect the PLH decoding performance. A solution to this problem has been proposed by designing PLH codes which allow for noncoherent detection, which can be successfully decoded even in presence of a phase offset. It is known that the performance of noncoherent detection is almost equivalent to those obtained with a coherent receiver, so the risk of incurring in a performance loss is negligible~\cite{CoRa00,CoRa01}.

Another example of the need of PLH codes allowing noncoherent detection has arisen from recent studies showing that precoding can ensure significant performance gains in a multibeam satellite system~\cite{ArGiCiErClAnVa16,VaCaPe16,GuVa20}. In a precoded system, the precoding matrix is computed at the transmitter in order to minimize the interference from other co-frequency signals, but it is unknown at the receiver. Hence, the received signals can be affected by a constant unknown amplitude and phase, that cannot be estimated by  means of pilot symbols, which are not precoded. After successful noncoherent detection of the PLH, it is possible to estimate the amplitude and phase of the subsequent information block transmitted in that frame, which is precoded with the same precoding matrix.

Typically, PLH codes have short length, as the above mentioned $64$-bit codewords included in the DVB-S2X standard. The design of short codes robust to noncoherent detection has been object of several research works, see for example\cite{KnLe94,JiRi02,Ru05,KaMo19}. Generally, communication standards foresee the use of constant length PLH headers. However, it has been proved that significant gains can be obtained by adopting variable length codes for the PLH~\cite{KaMo16}. This, however, makes the decoding of the PLH field especially challenging because the receiver needs to estimate also the length of the received PLH. Furthermore, when the PLH has a variable length, the position of pilot symbols is not known at the receiver prior to PLH decoding, so pilots cannot be used for phase estimation, thus further emphasizing the need for noncoherent detection. The short codeword length, however, makes PLH decoding performance practically unaffected by the presence of phase noise because the phase offset on the PLH field can be considered constant.

In order to successfully adopt variable length PLH codes in a realistic scenario, it is necessary to design a decoding scheme able to ensure good performance when the PLH codeword length is unknown at the receiver. In this paper, we adopt the strategy proposed in~\cite{KaMo19} to design a variable length code suitable for noncoherent detection, and we apply it to the DVB-S2X case study. We associate to each of the DVB-S2X ModCods a PLH codeword, with the aim of reducing the length of the PLH field with respect to the standard. We then design two alternative decoding strategies to jointly estimate the length of the PLH field and to successfully decode the transmitted codeword, which is the main contribution of the paper. We select the DVB-S2X scenario as an example to demonstrate the proposed strategies, but the same framework can be easily applied to other communication systems with different characteristics.

The paper is organized as follows. In Section~\ref{sec:model} we provide a general system model, and in Section~\ref{sec:plhcode} we briefly describe the adopted code. Section~\ref{sec:plhdet} represents the core of the paper, as it describes the designed decoding strategies, whose effectiveness is proved by a set of numerical simulations reported in Section~\ref{sec:results}. Finally, Section~\ref{sec:concl} concludes the paper.

\section{System Model}\label{sec:model}
The system considers the transmission of a sequence of information symbols with associated a PLH field of length $L$, belonging to a finite set of lengths, as specified in Table~\ref{tab:target SNR}, where the first column represents the signal-to-noise ratio (SNR) convergence threshold of the LDPC code for each ModCod, as specified in the DVB-S2X standard~\cite{DVB-S2X}. The details of the selected code design will be reported in the next section, while the meaning of the third and fourth columns of Table~\ref{tab:target SNR} will be clear at the end of this section.
\begin{table}
	\caption{PLH code details.}
	\label{tab:target SNR}
	\begin{center}
		\begin{tabular}{|c|c|c|c|}
			\hline
			SNR range [dB] & PLH Length $L$  & $N_L$ & $P(L)$  \\ 
			\hline \hline
			-2.85 & 64 & 1 & 0.0256 \\ 
			\hline 
			-2.03 & 58  & 1 & 0.0256 \\ 
			\hline 
			[0.22, 4.73] & 48 & 3 & 0.0769 \\ 
			\hline 
			[5.13, 19.57] & 26 & 34 & 0.8718 \\ 
			\hline 
		\end{tabular}
	\end{center}
\end{table}

Let us define by $L_{\mathrm{max}}$ the maximum value that $L$ can assume. It is always possible to consider the transmission of a vector of symbols of length $L_{\mathrm{max}}$, regardless of the true value of $L$. 
The vectors can be thus represented as divided in two distinct parts: the first one, indicated with subscript 1, of length $L$, contains the PLH symbols, while the second one, indicated with subscript 2, of length $\Delta L = L_{\mathrm{max}} - L$, contains unknown information symbols, which are drawn from a constellation with cardinality $M$, which depends on the selected ModCod.

Let us assume to transmit a vector of symbols $\boldsymbol{c}=[\boldsymbol{c}_1,\boldsymbol{c}_2]^{T}$ with length $L_{\mathrm{max}}$ such that $\boldsymbol{c}_1$ contains the PLH symbols and $\boldsymbol{c}_2$ contains randomly generated symbols of an unknown constellation.
We can define the received vector $\boldsymbol{r}$, of length $L_{\mathrm{max}}$, as
\begin{equation}\nonumber
	\boldsymbol{r}=A\boldsymbol{c}e^{j\theta} + \boldsymbol{w}
\end{equation}
where $\theta$ is an unknown channel phase, assumed constant and uniformly distributed in $[0,2\pi)$, $A$ is the unknown amplitude and $\boldsymbol{w}$ is a vector of i.i.d. complex Gaussian noise samples, whose real and imaginary components have variance $\sigma^2$. This is the scenario specified in the DVB-S2X implementation guidelines for broadcast applications~\cite{DVB-S2X_guide}.

All LDPC ModCods are considered equally likely. Since the number of ModCods associated to an arbitrary length $L$ is, in general, different, the probability to choose a PLH of length $L$ can take different values.
Assuming that the number of ModCods is $N$, the probability to select a PLH of length $L$ is 
\begin{equation}\nonumber
	P(L)=\frac{N_L}{N},
\end{equation}
where $N_L$ is the number of possible ModCods with PLH of length $L$. For the DVB-S2X standard and the adopted PLH code, described in the next section, the values of $N_L$ and $P(L)$ are reported in Table~\ref{tab:target SNR}, while $N=39$.

\section{Adopted PLH Code}\label{sec:plhcode}

We used the method presented in~\cite{KaMo19} for generating short codes that are robust to noncoherent detection for BPSK. We generate codes with length $V$, code size $K$ and minimum distance $d_\mathrm{{min}}$. The main idea behind the proposed algorithm is to add, at each step, a new column to the generator matrix $\mathbf{G}$, with dimension $K\times V$, such that the minimum distance $d_\mathrm{{min}}$ of the resulting new code is increased as much as possible. At each step, the column added to the generator matrix is checked to be different from all the existing columns.
The algorithm stops when a target $d_{\mathrm{min}}$ is achieved.

We associate to each of the DVB-S2X ModCods a PLH codeword whose length is reported in the second column of Table~\ref{tab:target SNR}, with the aim of reducing the length of the PLH field with respect to the standard.  We try to reduce as much as possible the code length by selecting the code with minimal length achieving the desired decision error rate at a target SNR.

In the design problem for PLH, given the desided error probability and the channel SNR, we need to construct the code with the shortest possible length. Therefore, a relation between these parameters is needed in order to correctly design the code. It can be demonstrated (see for example~\cite{CoRa00}) that a conservative upper bound for the codeword error probability $P_E$ when using noncoherent detection is given by
\begin{equation}\nonumber
	P_E\le\left(2^K-1\right)\frac{1}{2}\mathrm{erfc}\left(\sqrt{Bd_{\mathrm{min}}\frac{E_{\mathrm{s}}}{N_0}}\right)\,,
\end{equation} 
with $B=1$ and $B=1/2$ for BPSK and QPSK, respectively. This bound provides a relation between the desired codeword error probability, minimum distance, and SNR of the channel.

In our considered scenario, we need $K=6$ bits to distinguish among all the codewords. Assuming a target codeword error rate (CER) equal to $10^{-5}$ and considering that, in the worst case condition, $\frac{E_\mathrm{s}}{N_0}=-2.85$ dB (see~\cite{DVB-S2X}), we can obtain a lower bound for the minimum distance equal to $d_{\mathrm{min}}=26$. The corresponding PLH code with $K=6$ and $d_{\mathrm{min}}=26$ has length $V=64$~\cite{KaMo19}. This corresponds to the longest codeword in Table~\ref{tab:target SNR}. By repeating the procedure with the other target SNRs, we can obtain the other values of $L$ in the table.

\section{Designed PLH Decoding Strategies}\label{sec:plhdet}
In this section, we propose two alternative strategies for the decoding of a variable length PLH code as the one described in Section~\ref{sec:plhcode}.

The decoding strategy, when both the PLH and its length are unknown, can be based on the generalized likelihood criterion
\begin{equation}\label{likelihood criterion}
	\left(\hat{A},\hat{L},\hat{\boldsymbol{c}}\right)
	=\argmax_{(A,L,\boldsymbol{c})}P(\boldsymbol{c},L,A|\boldsymbol{r})\,.
\end{equation}
Equation~\eqref{likelihood criterion} gives an estimation of both the transmitted PLH codeword and its length. It also provides an estimation of the amplitude $A$. Clearly, it is 
\begin{equation}\nonumber
	P(\boldsymbol{c}|L)=
	\begin{cases}
	\frac{1}{N_L} & \text{if $\boldsymbol{c}$ is a codeword of length } L \\
	0                         & \text{otherwise.}
	\end{cases}
\end{equation}
Hence, with $\boldsymbol{c}$ a codeword of length $L$,
\begin{equation}\nonumber
	P(\boldsymbol{c}|L)P(L)=\frac{1}{N_L}\frac{N_L}{N}=\frac{1}{N}
\end{equation}
and, because $N$ does not depend on $L$, the decoding strategy in~\eqref{likelihood criterion} simplifies to
\begin{equation}\nonumber
	\left(\hat{A},\hat{L},\hat{\boldsymbol{c}}\right)
	=\argmax_{(A,L,\boldsymbol{c})}\frac{p(\boldsymbol{r}|\boldsymbol{c},L,A)}{p(\boldsymbol{r})}\,.
\end{equation}

Since we are considering a vector of length $L_{\mathrm{max}}$, thus independent of $L$, $p(\boldsymbol{r})$ is irrelevant in the maximization and it can be ignored. 
The generalized likelihood function can be rewritten as
\begin{equation}\label{likelihood_1}
	\begin{split}
		\left(\hat{A},\hat{L},\hat{\boldsymbol{c}}_1\right)&=\argmax_{(A,L,\boldsymbol{c}_1)}P(\boldsymbol{c}_1,L,A|\boldsymbol{r})
		\\&=\argmax_{(A,L,\boldsymbol{c}_1)}p(\boldsymbol{r}|\boldsymbol{c}_1,L,A)P(L)
		\\&=\argmax_{(A,L,\boldsymbol{c}_1)}p(\boldsymbol{r}|\boldsymbol{c}_1,L,A)\,.
	\end{split}
\end{equation} 

By exploiting the fact that vector $\boldsymbol{r}$ can be divided in two parts, the last probability density function (PDF) in~\eqref{likelihood_1} can be expressed as
\begin{equation}\label{likelihood_2}
	\begin{split}
		&p(\boldsymbol{r}|\boldsymbol{c}_1,L,A)
		=p(\boldsymbol{r}_1,\boldsymbol{r}_2|\boldsymbol{c}_1,L,A)
		\\&=\left(\frac{1}{M}\right)^{\Delta L}\sum_{\boldsymbol{c}_2}p(\boldsymbol{r}_1,\boldsymbol{r}_2|\boldsymbol{c}_1,\boldsymbol{c}_2,L,A)
		\\&=\left(\frac{1}{M}\right)^{\Delta L}\frac{1}{2\pi}\int\left[\sum_{\boldsymbol{c}_2}p(\boldsymbol{r}_1,\boldsymbol{r}_2|\boldsymbol{c}_1,\boldsymbol{c}_2,L,A,\theta)\right]\mathrm{d}\theta\,.
	\end{split}
\end{equation} 
Note that, in~\eqref{likelihood_2}, the average with respect to $\boldsymbol{c}_2$, that is necessary to obtain the marginal PDF in~\eqref{likelihood_1}, and the average w.r.t. the unknown phase $\theta$, are exploited. Given the fact that $\boldsymbol{r}_1$ and $\boldsymbol{r}_2$ are uncorrelated, it is possible to write
\begin{equation}\label{PDF}
	p(\boldsymbol{r}_1,\boldsymbol{r}_2|\boldsymbol{c}_1,\boldsymbol{c}_2,L,A,\theta)=
	p(\boldsymbol{r}_1|\boldsymbol{c}_1,L,A,\theta)
	p(\boldsymbol{r}_2|\boldsymbol{c}_2,L,A,\theta)\,.
\end{equation}
Hence, the two PDFs in~\eqref{PDF} can be computed separately. For the first one, we can write
\begin{equation}\label{eq:pr1_th}
	\begin{split}
		&p(\boldsymbol{r}_1|\boldsymbol{c}_1,L,A,\theta)
		=\prod_{k=1}^{L}p(r_k|c_k,A,\theta)
		\\&=\prod_{k=1}^{L} \left(\frac{1}{2\pi\sigma^2}\right) \exp\left\{-\frac{1}{2\sigma^2}|r_k|^2\right\}
		 \exp\left\{-\frac{A^2}{2\sigma^2}|c_k|^2\right\}
		 \\&\exp\left\{\frac{A}{\sigma^2}\Re\left[r_kc_k^*e^{-j\theta}\right]\right\}\,.
	\end{split}
\end{equation} 
As we will demonstrate in the next two subsections, because of some of the approximations involved in the derivation of the algorithms, the dependence on the channel phase $\theta$ will only appear in the first part of the right hand side of~\eqref{PDF}. Hence, we can average~\eqref{eq:pr1_th} over $\theta$ to obtain
\begin{equation}\label{eq:pr1}
	\begin{split}
		&p(\boldsymbol{r}_1|\boldsymbol{c}_1,L,A)
		=\frac{1}{2\pi}\int_{0}^{2\pi}p(\boldsymbol{r}_1|\boldsymbol{c}_1,L,A,\theta)\mathrm{d}\theta
		\\&=\left(\frac{1}{2\pi\sigma^2}\right)^L \exp\left\{-\frac{1}{2\sigma^2}\sum_{k=1}^{L}|r_k|^2\right\}
		 \\&\exp\left\{-\frac{A^2}{2\sigma^2}\sum_{k=1}^{L}|c_k|^2\right\}
		 \mathrm{I}_0\left(\frac{A}{\sigma^2}\left|\sum_{k=1}^{L}r_kc_k^*\right|\right)\,,
	\end{split}
\end{equation} 
in which $\mathrm{I}_0(\cdot)$ is the modified Bessel function of the first kind of order zero. 
For the second term in~\eqref{PDF}, instead, we can write
\begin{equation}\label{second term of PDF}
	p(\boldsymbol{r}_2|L,A,\theta)=
	\prod_{k=L+1}^{L_{\mathrm{max}}}\left[\frac{1}{M}\sum_{c_k}p(r_k|c_k,A,\theta)\right]\,.
\end{equation}
The term in square brackets in~\eqref{second term of PDF} cannot be computed in closed form, so some approximations are needed. In the following two subsections, we derive two alternative decoding rules based on different approximations.

\subsection{Strategy 1}\label{subsec:strategy1}
Let us consider one term of the summation in~\eqref{second term of PDF}. It can be written as
\begin{equation}\nonumber
	\begin{split}
		&p(r_k|c_k,A,\theta)
		=\left(\frac{1}{2\pi\sigma^2}\right)
		\exp\left\{-\frac{1}{2\sigma^2}|r_k|^2\right\}\\&
		\exp\left\{-\frac{A^2}{2\sigma^2}|c_k|^2+\frac{A}{\sigma^2}\Re\left[r_kc_k^*e^{-j\theta}\right]\right\}\,.
	\end{split}
\end{equation}
By using a first order Taylor approximation for the second exponential, we get
\begin{equation}\nonumber
	\begin{split}
		&p(r_k|c_k,A,\theta)
		\simeq\left(\frac{1}{2\pi\sigma^2}\right)
		\exp\left\{-\frac{1}{2\sigma^2}|r_k|^2\right\}\\&
		\left[1-\frac{A^2}{2\sigma^2}|c_k|^2+\frac{A}{\sigma^2}\Re\left[r_kc_k^*e^{-j\theta}\right]\right]\,.
	\end{split}
\end{equation}
Moreover, by averaging out terms with zero mean, we obtain that
\begin{equation}\nonumber
	\begin{split}
		\frac{1}{M}\sum_{c_k}p(r_k|c_k,A,\theta)\\
		&\hspace{-1.5cm}\simeq\left(\frac{1}{2\pi\sigma^2}\right)
		\exp\left\{-\frac{1}{2\sigma^2}|r_k|^2\right\}
		\left[1-\frac{A^2}{2\sigma^2}\right]
	\end{split}
\end{equation}
and, hence, we can write
\begin{equation}\label{eq:pr2_1}
	\begin{split}
		p(\boldsymbol{r}_2|L,A,\theta)\simeq\left(\frac{1}{2\pi\sigma^2}\right)^{\Delta L}\\
		&\hspace{-2.3cm}\exp\left\{-\frac{1}{2\sigma^2}\sum_{k=L+1}^{L_{\mathrm{max}}}|r_k|^2\right\}
		\left[1-\frac{A^2}{2\sigma^2}\right]^{\Delta L}\,,
	\end{split}
\end{equation}
where we notice that the dependence on $\theta$ has disappeared.
Finally, we can compute~\eqref{likelihood_2} by combining~\eqref{eq:pr1} and~\eqref{eq:pr2_1}:
\begin{equation}\nonumber
	\begin{split}
		&p(\boldsymbol{r}|\boldsymbol{c}_1,L,A)
		=\left(\frac{1}{2\pi\sigma^2}\right)^{L_{\mathrm{max}}}
		\exp\left\{-\frac{1}{2\sigma^2}\sum_{k=1}^{L_{\mathrm{max}}}|r_k|^2\right\}\\&
		\exp\left\{-\frac{A^2}{2\sigma^2}\sum_{k=1}^{L}|c_k|^2\right\}
		\mathrm{I}_0\left(\frac{A}{\sigma^2}\left|\sum_{k=1}^{L}r_kc_k^{*}\right|\right)
		\left[1-\frac{A^2}{2\sigma^2}\right]^{\Delta L}\,.
	\end{split}
\end{equation}
Being $L_{\mathrm{max}}$ constant, the first two terms are irrelevant.
For the last one we assume to make the following approximation, which is good if $\sigma^2\gg1$:
\begin{equation}\nonumber
	\left[1-\frac{A^2}{2\sigma^2}\right]^{\Delta L}\simeq 1\,.
\end{equation}
Hence, by exploiting the approximation $\mathrm{I}_0(x)\simeq e^x$ and taking the logarithm, the decision rule becomes
\begin{equation}\label{final_likelihood}
	\left(\hat{A},\hat{L},\hat{\boldsymbol{c}}_1\right)
	=\argmax_{(A,L,\boldsymbol{c}_1)}\left(A\left|\sum_{k=1}^{L}r_kc_k^{*}\right|
	- \frac{A^2L}{2}\right)\,.
\end{equation} 
By performing the maximization with respect to $A$ first, we obtain that the amplitude estimate is
\begin{equation}\label{Amplitude_estimate}
	\hat{A}=\frac{\left|\sum_{k=1}^{L}r_kc_k^{*}\right|}{L}\,.
\end{equation} 
Substituting~\eqref{Amplitude_estimate} in~\eqref{final_likelihood}, we have
\begin{equation}
	\left(\hat{L},\hat{\boldsymbol{c}}_1\right)
	=\argmax_{(L,\boldsymbol{c}_1)}
	\frac{\left|\sum_{k=1}^{L}r_kc_k^{*}\right|^2}{L}\,.
\end{equation} 
Considering that we used several approximations in the derivation of this decision rule, some of them involving the PLH length, it can be convenient to correct the metric as follows
\begin{equation}\label{metrica finale beta}
	\left(\hat{L},\hat{\boldsymbol{c}}_1\right)
	=\argmax_{(L,\boldsymbol{c}_1)}
	\frac{\left|\sum_{k=1}^{L}r_kc_k^{*}\right|^2}{L^{\alpha}}\,,
\end{equation} 
where $\alpha$ is a coefficient to be optimized.

\subsection{Strategy 2}\label{subsec:strategy2}
An alternative decoding scheme can be derived by considering the following approximation
\begin{equation}\nonumber
	c_k\simeq E\left\{|c_k|\right\}e^{j\psi_k}=e^{j\psi_k}\,,
\end{equation}
where $\{\psi_k\}$ are independent random variables with uniform distribution in $[0,2\pi)$.
By averaging with respect to $\psi_k$, we can rewrite~\eqref{second term of PDF} as
\begin{equation}\nonumber
	\begin{split}
		&p(\boldsymbol{r}_2|L,A,\theta)\\
		&=\prod_{k=L+1}^{L_{\mathrm{max}}}\frac{1}{2\pi}\int_{0}^{2\pi}\left[\frac{1}{M}\sum_{c_k}p(r_k|c_k,A,\theta)\right]\mathrm{d}\psi_k\\
		&=\left(\frac{1}{2\pi\sigma^2}\right)^{\Delta L}
		\exp\left\{-\frac{1}{2\sigma^2}\sum_{k=L+1}^{L_{\mathrm{max}}}|r_k|^2\right\}
		\exp\left\{-\frac{A^2}{2\sigma^2}\Delta L\right\}\\&
		\prod_{k=L+1}^{L_{\mathrm{max}}}\frac{1}{2\pi}\int_{0}^{2\pi}\exp\left\{\frac{A}{\sigma^2}\Re\left[r_k e^{-j\psi_k} e^{-j\theta} \right]\right\}\mathrm{d}\psi_k\\
		&=\left(\frac{1}{2\pi\sigma^2}\right)^{\Delta L}
		\exp\left\{-\frac{1}{2\sigma^2}\sum_{k=L+1}^{L_{\mathrm{max}}}|r_k|^2\right\}
		\exp\left\{-\frac{A^2}{2\sigma^2}\Delta L\right\}\\&
		\prod_{k=L+1}^{L_{\mathrm{max}}}\frac{1}{2\pi}\int_{0}^{2\pi}\exp\left\{\frac{A}{\sigma^2}|r_k|\cos\left(\psi_k+\theta-\angle{r_k} \right)\right\}\mathrm{d}\psi_k\,.
\end{split}
\end{equation}
Using second order Taylor approximation for the last exponential, we obtain
\begin{equation}\nonumber
	\begin{split}
		&p(\boldsymbol{r}_2|L,A,\theta)\\
		&\simeq\left(\frac{1}{2\pi\sigma^2}\right)^{\Delta L}
		\exp\left\{-\frac{1}{2\sigma^2}\sum_{k=L+1}^{L_{\mathrm{max}}}|r_k|^2\right\}
		\exp\left\{-\frac{A^2}{2\sigma^2}\Delta L\right\}\\
		&\prod_{k=L+1}^{L_{\mathrm{max}}}\frac{1}{2\pi}\int_{0}^{2\pi}\left[1+\frac{1}{\sigma^2}|r_k|\cos\left(\psi_k+\theta-\angle{r_k}\right)\right.\\
		& \left.+ \frac{1}{2\sigma^4}|r_k|^2\cos^2\left(\psi_k+\theta-\angle{r_k}\right)\right]^A\mathrm{d}\psi_k\,.
\end{split}
\end{equation}
Adopting now the approximation $(1+x)^r\simeq 1+rx$ for $|x|\ll1$, we get
\begin{equation}\label{eq:pr2_2}
	\begin{split}
		&p(\boldsymbol{r}_2|L,A,\theta)\\
		&\simeq\left(\frac{1}{2\pi\sigma^2}\right)^{\Delta L}
		\exp\left\{-\frac{1}{2\sigma^2}\sum_{k=L+1}^{L_{\mathrm{max}}}|r_k|^2\right\}
		\exp\left\{-\frac{A^2}{2\sigma^2}\Delta L\right\}\\
		&\prod_{k=L+1}^{L_{\mathrm{max}}}\frac{1}{2\pi}\int_{0}^{2\pi}\left[1+\frac{A}{\sigma^2}|r_k|\cos\left(\psi_k+\theta-\angle{r_k}\right)\right.\\
		& \left.+ \frac{A}{2\sigma^4}|r_k|^2\cos^2\left(\psi_k+\theta-\angle{r_k}\right)\right]\mathrm{d}\psi_k\\
		&=\left(\frac{1}{2\pi\sigma^2}\right)^{\Delta L}
		\exp\left\{-\frac{1}{2\sigma^2}\sum_{k=L+1}^{L_{\mathrm{max}}}|r_k|^2\right\}
		\exp\left\{-\frac{A^2}{2\sigma^2}\Delta L\right\}\\
		&\prod_{k=L+1}^{L_{\mathrm{max}}} \left(1+ \frac{A}{8\pi\sigma^4}|r_k|^2\right)\,.
	\end{split}
\end{equation}
Note that this expression is independent of $\theta$. Hence, we can compute~\eqref{likelihood_2} by using~\eqref{eq:pr1} and~\eqref{eq:pr2_2} as
\begin{equation}\nonumber
\begin{split}&
p(\boldsymbol{r}|\boldsymbol{c}_1,L,A)
=\left(\frac{1}{2\pi\sigma^2}\right)^{L_{\mathrm{max}}}
\exp\left\{-\frac{1}{2\sigma^2}\sum_{k=1}^{L_{\mathrm{max}}}|r_k|^2\right\}\\&
\mathrm{I}_0\left(\frac{A}{\sigma^2}\left|\sum_{k=1}^{L}r_kc_k^*\right|\right)
\exp\left\{-\frac{A^2}{2\sigma^2}L_{\mathrm{max}}\right\}\\&
\prod_{k=L+1}^{L_{\mathrm{max}}} \left(1+ \frac{A}{8\pi\sigma^4}|r_k|^2\right)\,.
\end{split}
\end{equation}
By removing the first two irrelevant terms, applying the approximation $\mathrm{I}_0(x)\simeq e^x$, and taking the logarithm, the likelihood function can be written as
\begin{equation}\nonumber
	\begin{split}
		&\left(\hat{A},\hat{L},\hat{\boldsymbol{c}}_1\right)
		=\argmax_{(A,L,\boldsymbol{c}_1)}\left(\frac{A}{\sigma^2}\left|\sum_{k=1}^{L}r_kc_k^{*}\right|\right.\\
		& \left. + \sum_{k=L+1}^{L_{\mathrm{max}}}\log \left(1+ \frac{A}{8\pi\sigma^4}|r_k|^2\right)
		- \frac{A^2L_\mathrm{max}}{2\sigma^2}\right)\,.
	\end{split}
\end{equation} 
Approximating $\log(1+x)\simeq x$ for $|x|\ll 1$, we get
\begin{equation}\label{eq:str2_A}
	\begin{split}
		&\left(\hat{A},\hat{L},\hat{\boldsymbol{c}}_1\right)
		=\argmax_{(A,L,\boldsymbol{c}_1)}\left(A\left|\sum_{k=1}^{L}r_kc_k^{*}\right|\right.\\
		& \left. + \frac{A}{8\pi\sigma^2}\sum_{k=L+1}^{L_{\mathrm{max}}}|r_k|^2
		- \frac{A^2L_\mathrm{max}}{2}\right)\,.
	\end{split}
\end{equation}
Also in this case, we perform the maximization with respect to $A$ first to obtain the following amplitude estimate value,
\begin{equation}\label{Amplitude_estimate2_2}
	\hat{A}=\frac{\left|\sum_{k=1}^{L}r_kc_k^{*}\right|+\frac{1}{8\pi\sigma^2}\sum_{k=L+1}^{L_{\mathrm{max}}}|r_k|^2}{L_\mathrm{max}}\,.
\end{equation} 
Finally, by replacing~\eqref{Amplitude_estimate2_2} in~\eqref{eq:str2_A}, we obtain the decision rule
\begin{equation}\nonumber
	\begin{split}
	\left(\hat{L},\hat{\boldsymbol{c}}_1\right)
	&=\argmax_{(L,\boldsymbol{c}_1)}
	\left(\left|\sum_{k=1}^{L}r_kc_k^{*}\right|+\frac{1}{8\pi\sigma^2}\sum_{k=L+1}^{L_{\mathrm{max}}}|r_k|^2\right)^2\\
	&=\argmax_{(L,\boldsymbol{c}_1)}
	\left|\sum_{k=1}^{L}r_kc_k^{*}\right|+\frac{1}{8\pi\sigma^2}\sum_{k=L+1}^{L_{\mathrm{max}}}|r_k|^2\,.
	\end{split}
\end{equation} 
Also in this case, we can apply a correction factor $\beta$ to the metric, to be optimized, in order to take into account the approximations involved in the derivation. Hence, the final metric is
\begin{equation}\label{metrica finale alpha}
	\left(\hat{L},\hat{\boldsymbol{c}}_1\right)
	=\argmax_{(L,\boldsymbol{c}_1)}
	\left|\sum_{k=1}^{L}r_kc_k^*\right| +
	\frac{\beta}{\sigma^2}\sum_{k=L+1}^{L_{\mathrm{max}}}|r_k|^2\,.
\end{equation} 
We can notice that this strategy requires knowledge of the noise variance $\sigma^2$.

\section{Numerical Results}\label{sec:results}
In this section, we show the performance of the proposed decoding schemes derived in Sections~\ref{subsec:strategy1} and~\ref{subsec:strategy2} in the considered DVB-S2X scenario.
For each possible length of the PLH, we select the ModCod associated with the lowest SNR, in order to test the decoder in the worst working conditions. This corresponds to ModCods 1, 2, 3, and 6, as can be observed from Table~\ref{tab:target SNR}.
The performance is compared with that achievable with the code adopted by the DVB-S2X standard~\cite{DVB-S2X}. We remind that the standard uses a fixed-length $64$-bits PLH code, designed to guarantee reliable performance in all working conditions. A receiver for the standard code operates according to the following decoding rule
\begin{equation}\label{eq:ideal_dec}
\boldsymbol{\hat{c}} = 
\argmax_{\boldsymbol{c}}\left|\sum_{k=1}^{L_\mathrm{max}}r_kc_k^*\right|\,,
\end{equation}
and its performance will be labeled as \emph{standard} in this section.
We also compare the proposed decoding strategies with another benchmark strategy, denoted as \emph{simple} in this section. This decoder does not know the codeword length and it simply computes the likelihood function of each codeword with the received signal, assuming always the maximum length $L_{\mathrm{max}}$, and it operates according to decoding rule~\eqref{eq:ideal_dec}. This is representative of the performance achievable in a system adopting a variable length PLH, but using a conventional detection scheme.

\subsection{Strategy 1}
\begin{figure}
	\centering
	\includegraphics[width=1\columnwidth]{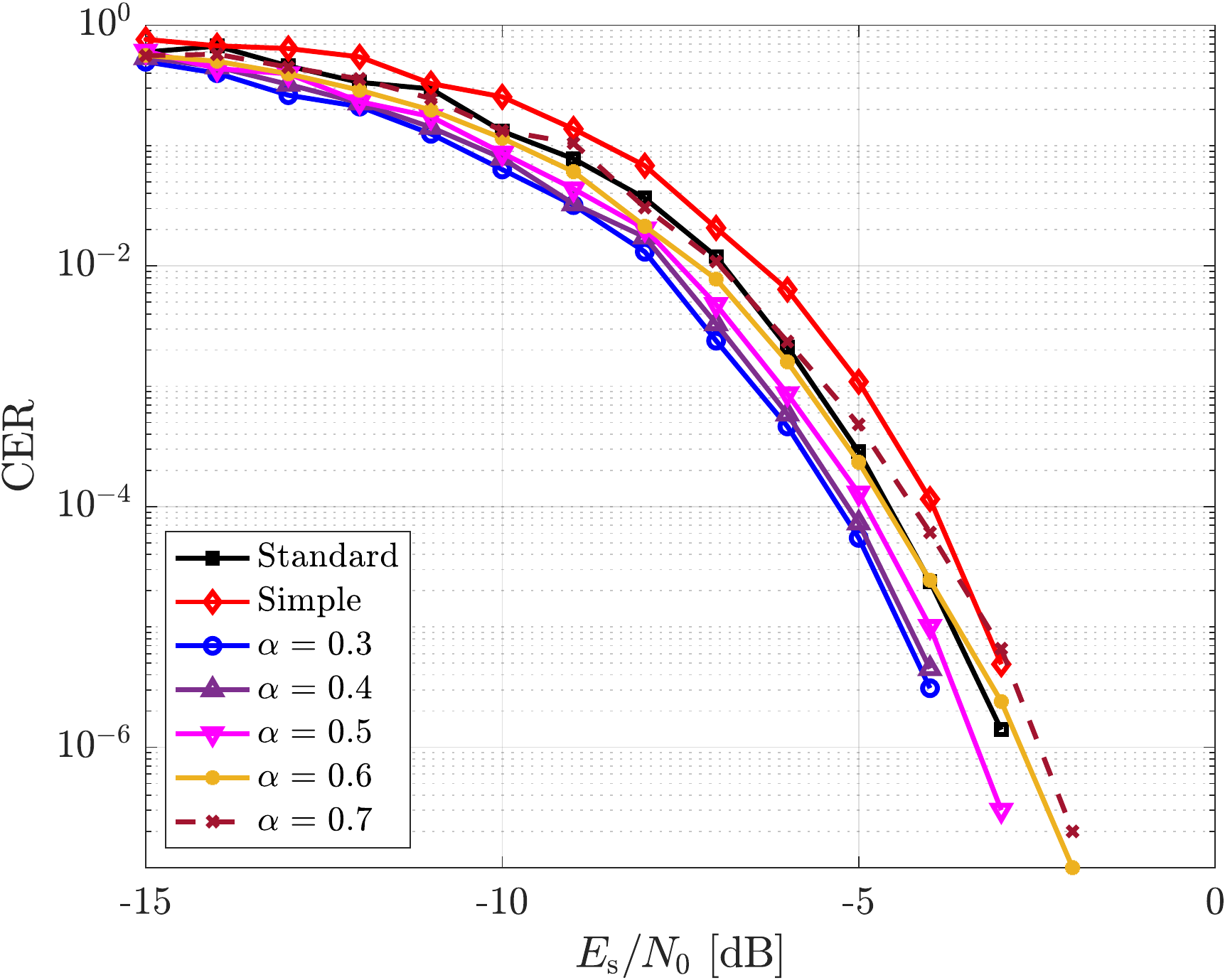}
	\caption{PLH CER for ModCod 1 using Strategy 1.}
	\label{fig:der1_1}
\end{figure}

\begin{figure}
	\centering
	\includegraphics[width=1\columnwidth]{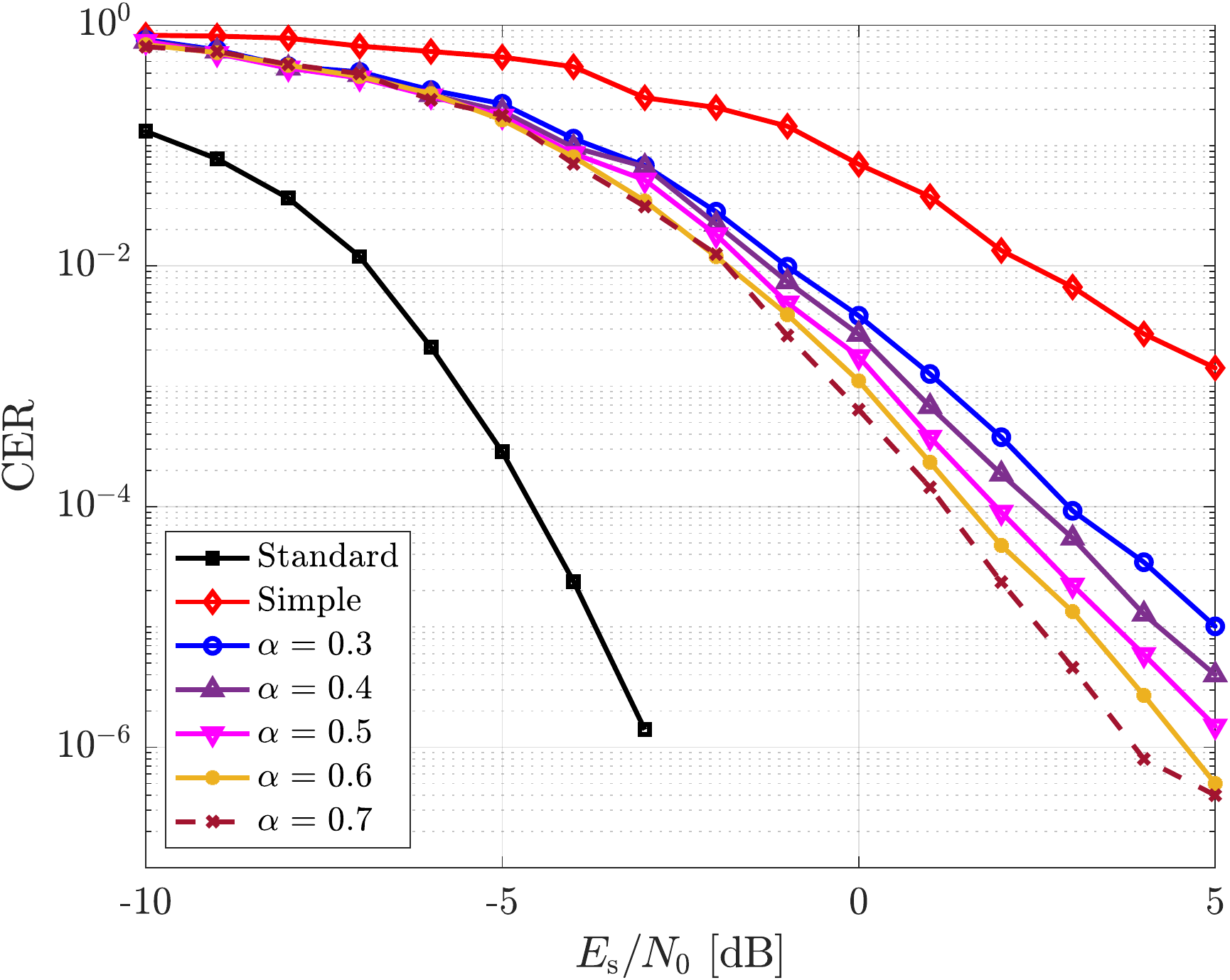}
	\caption{PLH CER for ModCod 6 using Strategy 1.}
	\label{fig:der6_1}
\end{figure}

The performance of the first proposed strategy is reported in Figs.~\ref{fig:der1_1} and~\ref{fig:der6_1} as a function of the ratio between the average energy per symbol and the noise power spectral density, $E_\mathrm{s}/N_0$, for ModCods 1 and 6, respectively. Different values of the parameter $\alpha$ in~\eqref{metrica finale beta} are compared. We can see that for ModCod $1$ all tested values of $\alpha$ provide relatively similar performance, and that increasing the value of $\alpha$ the performance tends to degrade. We also notice that, in this case, the standard and the simple receivers reach similar performance. This is not surprising, because in our proposed scheme ModCod 1 adopts a $64$-bits PLH codeword, which is the same length considered by both the standard and simple schemes.

If we look at ModCod 6, instead, first of all we notice the poor performance of the simple receiver scheme. Also in this case, this was expected, recalling that this scheme attempts decoding on $64$ bits, but the PLH codeword for ModCod 6 is only 26 bits long, so the decoder tries to use also part of the information symbols, which are encoded and modulated with the code and modulation specific to the ModCod. We can also notice the large gap between the standard decoder and our proposed solution. This is also unsurprising, considering that the standard code needs to work properly for all ModCods, including those operating at low SNR, so it has to be robust to the worst case scenario. Adopting a variable length code, and a corresponding decoding strategy, allows us to relax this robustness constraint for ModCods operating at higher SNR, like ModCod 6 and those above. If we compare the curves for different values of $\alpha$, we see that the performance improves by increasing $\alpha$. This is exactly the opposite of the behavior we observed for ModCod 1, so the problem arises of how to select $\alpha$ to guarantee a good trade-off in all cases.

\begin{figure}
	\centering
	\includegraphics[width=1\columnwidth]{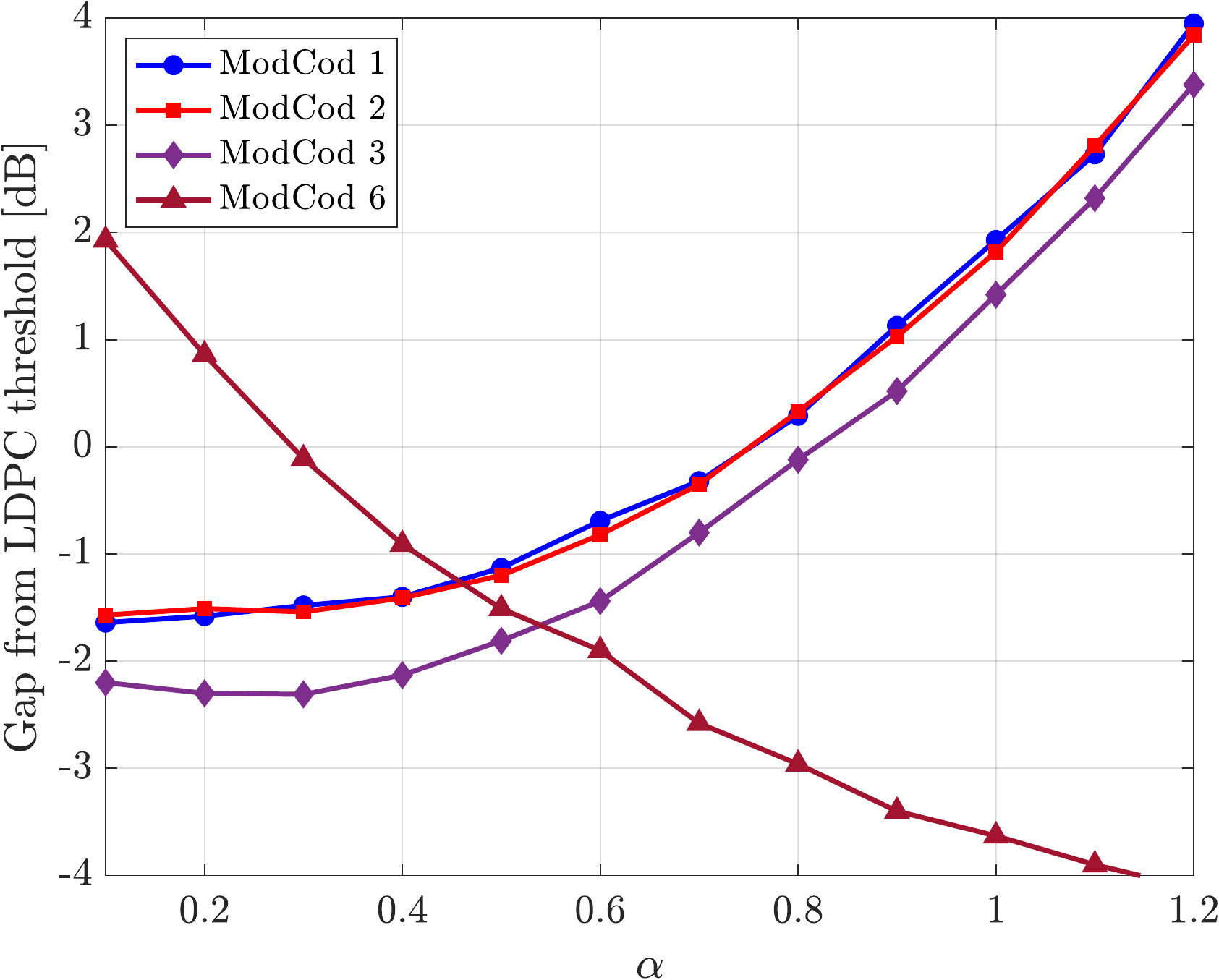}
	\caption{Gap in dB from the ModCod convergence threshold for several values of $\alpha$ at ${\mathrm{CER}=10^{-5}}$, using Strategy 1.}
	\label{fig:gap-from-modcod-threshold-at-1e-5}
\end{figure}

In order to allow for a correct decoding of the transmitted information, it is important that the PLH is reliably decoded at SNR values that are lower than the working conditions of the ModCods.
Fig.~\ref{fig:gap-from-modcod-threshold-at-1e-5} shows the gap from the ModCod convergence threshold, as reported in~\cite{DVB-S2X}, for several values of $\alpha$, at a target ${\mathrm{CER}=10^{-5}}$. 
From the figure, we can notice that the maximum gap is minimized for $\alpha=0.5$. 
Moreover, we can see that for $0.3\le\alpha\le0.7$ the gaps are negative, which means that the PLH is decoded with a CER below $10^{-5}$ for SNR values which are lower than the ModCod convergence threshold. This is fundamental to guarantee the correct decoding of all ModCods.
Considering the probability of occurrence of the different PLH lengths, we notice that choosing $\alpha=0.7$ is a value which mostly favors the most probable length and still keeps the PLH decoding performance below the LDPC convergence threshold for all ModCods.
Values of $\alpha$ higher than $0.7$ will reduce the gap for ModCod $6$ but also increase the gap for the other ModCods, so they should not be considered.

\subsection{Strategy 2}
The results obtained using Strategy 2 in Section~\ref{subsec:strategy2} are presented in this section.
Through simulation, we verified that the metric shows the best performance for $\beta=0.2$ for all ModCods; the gaps from the ModCods convergence threshold are presented in Table~\ref{tab:gap-alpha}. 
\begin{table}
	\caption{Gap in dB from the ModCod convergence threshold at ${\mathrm{CER}=10^{-5}}$ using Strategy 2 with $\beta=0.2$.}
	\begin{center}
		\begin{tabular}{|c|c|}
			\hline 
			ModCod & Gap from LDPC threshold \\ 
			\hline\hline
			1 &  -0.86 \\ 
			\hline 
			2 &  -1.04 \\ 
			\hline 
			3 &  -1.67 \\ 
			\hline 
			6 &  -4.35  \\ 
			\hline 
		\end{tabular} 
		\label{tab:gap-alpha}
	\end{center}
\end{table}
We see that for all the ModCods the PLH is decoded reliably at SNR values that are lower than the ModCods convergence threshold; in particular, for ModCod 6, a very large gap can be ensured.

\begin{figure}
	\centering
	\includegraphics[width=1\columnwidth]{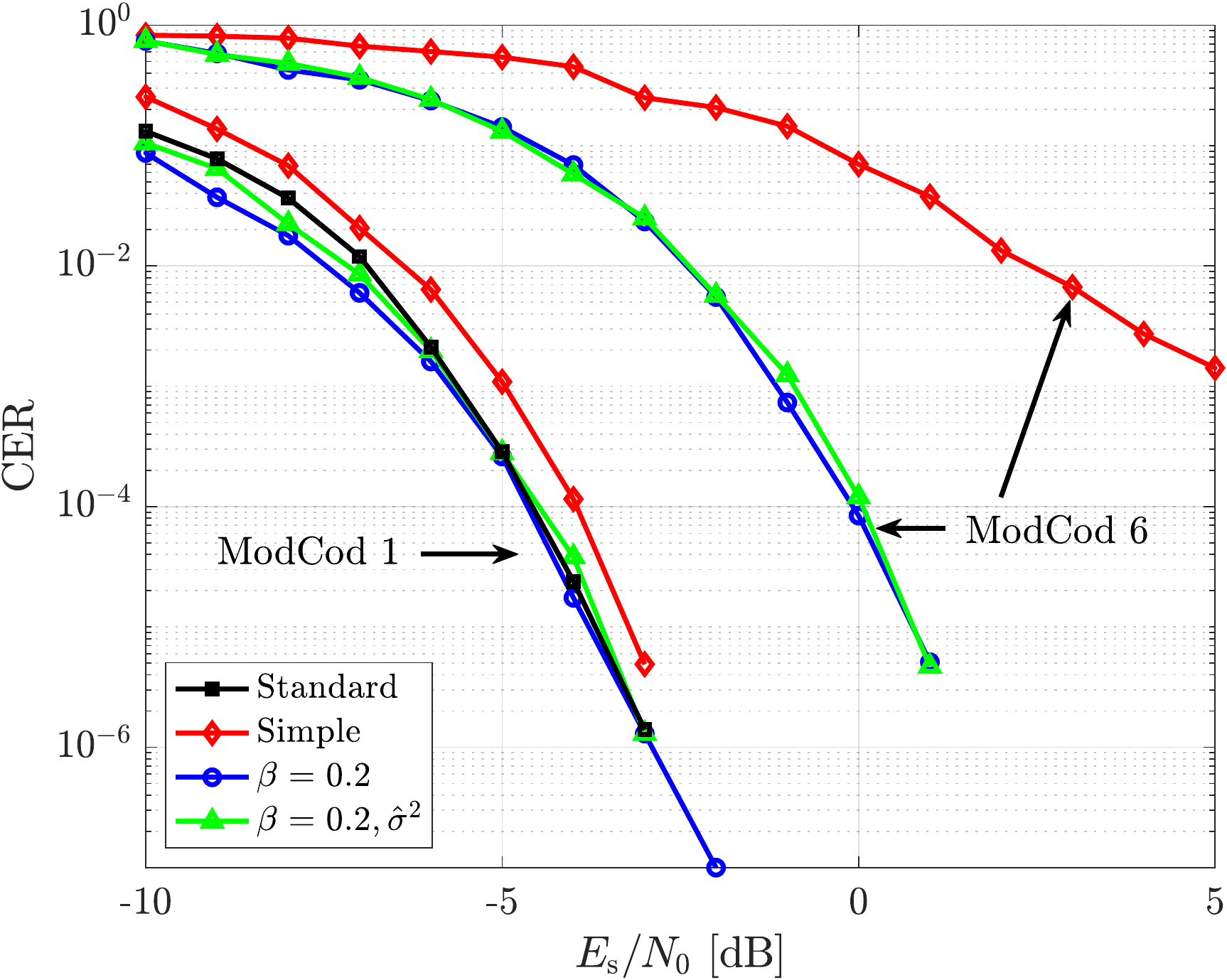}
	\caption{PLH CER comparison for Strategy 2 using a noise variance estimator.}
	\label{fig:der1_6_str2_var}
\end{figure}

As mentioned, Strategy 2 requires knowledge of the noise variance. The results reported in Table~\ref{tab:gap-alpha} assume perfect knowledge of the variance, but this parameter can be estimated~\cite{PaBe00}. We resort to the algorithm proposed in~\cite{KaJe07}, which is based on the second and fourth order moments of the received signal, and does not need pilot symbols to perform estimation, so it is independent of the adopted frame structure. Fig.~\ref{fig:der1_6_str2_var} reports the CER curves for ModCods 1 and 6 assuming ideal knowledge of the noise variance and using the variance estimator~\cite{KaJe07} ($\hat{\sigma}^2$ curves). We notice that we can achieve the same performance with the variance estimator as in case of perfect knowledge of the variance.

\subsection{Comparison and Discussion}
We have proposed two alternative decoding strategies, with different complexity and performance. Clearly, the best decoding strategy depends on the probability distribution of the PLH length over the ModCods, as well as on the number of ModCods and on the adopted PLH code. However, both proposed decoding strategies can ensure very good performance using a PLH with length significantly lower than that used in the DVB-S2X standard. In fact, if we compare the proposed design with the standard, where all $39$ ModCods adopt a fixed length $64$-bit code, and taking into account the probability distribution of the codeword length reported in Table~\ref{tab:target SNR}, we can notice that the average codeword length is $30$ bits, so less than half the value adopted in the standard.

If we compare Strategy 1 and Strategy 2, we notice that Strategy 2 has slightly better performance, in terms of larger gaps from LDPC convergence threshold, at the cost of an increased complexity. In fact, Strategy 1~\eqref{metrica finale beta} operates only on the first $L$ symbols, while Strategy 2~\eqref{metrica finale alpha} takes all $L_\mathrm{max}$ symbols into account. A standard fixed length PLH decoding strategy would operate with the metric~\eqref{eq:ideal_dec} on all $L_\mathrm{max}$ symbols, and the same would happen for the described simple receiver. Table~\ref{tab:compl} reports the complexity of the different strategies, in terms of number of real operations. To evaluate the complexity, we have to keep in mind that the PLH uses $\pi/2-$BPSK symbols, so multiplications with preamble symbols only results in a change of the sign of the real or the imaginary part. Hence, multiplications by preamble symbols are not accounted for. Look-up table (LUT) accesses are necessary to evaluate square roots to compute absolute values. For Strategy 1 and 2 we have used the average codeword length, $30$, to evaluate the complexity. We notice that Strategy 1 has a largely inferior complexity with respect to both Strategy 2 and a standard decoding scheme, while Strategy 2 is comparable to the standard scheme in terms of additions and it requires more multiplications. Hence, Strategy 1 is clearly desirable from an implementation complexity point of view. We have to keep in mind, however, that the performance and complexity may vary depending on the PLH length distribution, number of PLH codewords and the channel conditions. Hence, while the general decoding framework remains valid, performance must be evaluated according to the characteristics of the considered communication system.

\begin{table}
	\caption{Complexity comparison of the proposed decoding strategies.}
	\begin{center}
		\begin{tabular}{|c|c|c|c|}
		\hline
		Strategy   & \# additions         & \# multiplications   & \# LUT \\ \hline\hline
		Standard~\eqref{eq:ideal_dec}   & $2L_\mathrm{max}-1=127$              & $2$                     & $1$                \\ \hline
		Strategy 1~\eqref{metrica finale beta} & $2L-1=59$               & $2$                     & $0$                \\ \hline
		Strategy 2~\eqref{metrica finale alpha} & $2L_\mathrm{max}-4=124$ & $2L_\mathrm{max}-2L=68$ & $1$                \\ \hline
		\end{tabular}
		\label{tab:compl}
	\end{center}
\end{table}

\section{Conclusions}\label{sec:concl}
We proposed two decoding strategies to decode a PLH code with variable length codewords. The use of a variable length code for the PLH field allows for a significant reduction of the overhead in each of the transmitted frames, but makes decoding harder because the codeword length is unknown at the receiver. We took the DVB-S2X standard as a test case, and showed that excellent decoding performance can be achieved by estimating the unknown length of the PLH field, together with the decoding of the PLH code. The proposed decoding strategies ensure that a very low error rate can be reached at SNR values which allow for a successful decoding of the information enclosed in the frames. The proposed decoders rely on the tuning of a parameter, which takes into account the approximations involved in the derivation. The optimal value of this parameter, clearly, depends on the investigated scenario and on the adopted PLH code, but the general strategy can be easily applied to different cases and applications.

\bibliographystyle{ieeetr}

\end{document}